# Dose-dependent isotherm of Kr adsorption on heterogeneous bundles of closed single-walled carbon nanotubes

Svetlana Yu. Tsareva[*], Edward McRae, Fabrice Valsaque and Xavier Devaux

Institut Jean Lamour, UMR 7198 CNRS - Université de Lorraine,
Faculté des Sciences et Technologies, Boulevard des Aiguillettes, BP 70239,
54506 Vandœuvre-lès-Nancy, France

**ABSTRACT:** We present 77K isotherms of krypton adsorption on bundles of closed highly-pure HiPco single-walled carbon nanotubes (SWCNTs). Two volumetric adsorption protocols were used, one with an increasing Kr dose per injection (IAD), one with a constant dose (CAD). Detailed microstructural examination showed that the SWCNTs combine into small bundles (of 25-30 SWCNTs) which are heterogeneous in diameter with a consequential range of interstitial channel (IC) shapes and sizes, some of which are large. The IC-sites are the subnanoscaled pores with alternating enlargements and constrictions along the tube axes. This results in adsorption dosing (AD) dependent characteristics of the low-pressure region of the isotherm. In the IAD protocol an equilibrium pressure drop with increasing AD is noted. Using the CAD protocol, different branches are observed. Well-pronounced substeps were established which we interpret as corresponding to the formation of various phases of confined Kr with different atoms arrangement. The height of a given substep obtained in different measurements depends on the AD value which can strongly influence the population of the site. Some substeps existing only for certain values of AD suggests the existence of a certain selectivity or of a preferential phase formation according to this value.

**Keywords:** Single-walled carbon nanotubes bundles, Kr adsorption isotherms, High-resolution transmission electron microscopy, DRIFT spectroscopy

---

[*] Corresponding author: E-mail: svetlana.tsareva@univ-lorraine.fr

# 1. Introduction

Carbon nanotubes (CNTs), and especially single-walled carbon nanotube bundles, are remarkable adsorption substrates because of their unique structural properties. A single-walled carbon nanotube (SWCNT) bundle may be represented schematically as a quasi-one-dimensional structure with arrays of four different possible adsorption sites (Williams and Eklund 2000; Muris et al. 2001). These include the 1D sites located in the grooves (G) separating two adjacent outer tubes, the inner channels of CNTs (T) if they are open and the interstitial channels (IC). The external walls (E) of the outermost SWCNTs in a bundle are the sites of 2D adsorption.

T- and IC-sites are what are referred to in other fields as nanopore channels. The adsorption inside pores is defined by the size and the geometry of the channels (Maddox et al 1995; Horikawa et al. 2011; Sadeghi et al. 2012; Cimino et al. 2013; Puibasset 2014; Nguyen 2013). Whether endohedral adsorption inside open SWCNTs can occur is determined by the ratio of adsorbate to CNT diameter (Kuznetsova et al. 2000; Babaa et al. 2003; Jakubek and Simard 2004; Jakubek and Simard 2005; Arab et al. 2006). The size of IC-sites in closely-packed homogeneous-diameter nanotube bundles is smaller than that of the T-sites. In fact, they are subnanoscaled. Therefore, some authors have refuted the possibility of any adsorption in the IC-sites (Talapatra et al. 2000) for Ne, Xe and $CH_4$. On the other hand, Calbi et al. (2001) calculated that some gases can enter the IC-sites upon a coherent bundle dilation. They predicted that the lattice dilation for He and Ne would be less than 0.5%. In the case of $H_2$, Ar and $CH_4$ the expansion would be more significant and could reach ~2.5%. Filling of interstitial sites by He (Stan et al. 1998), Ne (Stan et al. 1998), $CH_4$ (Muris et al. 2000; Lohnson et al. 2003; Shi et al. 2003), $N_2$ (Fujiwara et al. 2001), $O_2$ (Fujiwara et al. 2001), Kr (Muris et al. 2000), Ar (Shi et al. 2003; Bienfait et al. 2003), Xe (Shi et al. 2003) has been reported. Neutron scattering measurements accompanying the adsorption experiments have shown that there is no dilation when Ar (Bienfait et al. 2003) or $CH_4$ (Johnson et al. 2003) is adsorbed in the bundle. Johnson et al. (2003) thus suggested that the filling of IC-sites is accompanied by tube deformation, or an ovalization. When methane is forced into IC-sites, the tubes deform but the nanotube carbon atoms furthest from the methane remain close to their original positions. Distortion of SWCNTs in a bundle has been proven in some works (Vodenitcharova et al. 2007; Jiang et al. 2008; Zang et al. 2009; López et al. 2001). For example, for bundles of two identical SWCNTs (1.56 and 2.05 nm) in diameter Jiang et al. (2008) determined that the flattening values could reach respectively 2.2% and 3.4% of the diameter. The deformation is a function of CNT diameter as was confirmed by a high-resolution transmission electron microscopy investigation (Abrams and Hanein 2007). So, the deformation of SWCNTs in the bundle might be an important factor in adsorption, both modifying the size of IC-sites and allowing the stabilization of the adsorbed molecules (Johnson et al. 2003).

For a given adsorbate, the accessibility of the IC-sites depends on the tube diameters. Arab et al. (2007) estimated for a perfect bundle formed by three (10,10) SWCNTs that the passage from repulsive to attractive energies is extremely rapid for tube radii around 0.7 and 1.0 nm for Kr and Xe, respectively. In reality, bundles are inhomogeneous due to the heterogeneity of the SWCNT diameter distribution. Bundles containing a distribution of nanotube diameters always give structures with packing defects that may form relatively large ICs (Shi et al. 2003) which consequently may be accessible to an adsorbate according to its size.

Rare gas adsorption isotherms of SWCNTs are more informative than those based on measurements with nitrogen since they allow distinguishing the 1D and 2D adsorption sites. Generally speaking, there are two inclined steps as the relative equilibrium pressure progressively increases (Muris et al. 2001; Babaa et al. 2003; Arab et al. 2006; Johnson et al. 2003). The lower pressure step represents adsorption on higher binding energy sites (G-, T- and IC-sites). The higher pressure step corresponds to the exohedral adsorption on the external surface of the bundles (E-sites).

A presence of impurities such as amorphous carbon, hollow carbon shells or carbon-covered catalyst particles complicates the interpretation of adsorption isotherms since they may also constitute adsorption sites. These can modify the shape of an isotherm by smoothing or even leading to the disappearance of the steps.

In this work we will present Kr isotherms on highly pure HiPco SWCNTs. We have found some unexpected results using our carefully developed protocol. We show that such a protocol leads to even richer results than those found to the present and may yield more detailed information than what is usual for SWCNTs in real bundles with different tube diameters. Through electron microscopy results we present features not only of the diameter distribution but also of the wide range of possible interstitial channel morphologies and dimensions. The results provoke many new questions and we will attempt to furnish some potential interpretations.

## 2. Experimental

### 2.1 Samples

The SWCNTs used in this study are HiPco SuperPureTubes™ produced by NanoIntegris. According to the manufacturer, the diameter range is 0.8-1.2 nm with individual CNT lengths up to 1 μm; they contain <5% of metal catalyst impurities. The SWCNTs are bundled. (website: http://www.nanointegris.com/en/downloads see "HiPco Technical Data Sheet")

A soft sample oxidation was made by treatment with $H_2SO_4$ (2.5M) under microwave irradiation for 20 min at 50°C. The SWCNT acid suspension was neutralized by adding a sodium hydroxide solution (2M, 4 mL). The ox-SWCNTs were washed extensively with water using sonication. The final rinsing was performed with ethanol.

### 2.2 Volumetric measurements

The physisorption studies were carried out using an apparatus with two capacitive pressure gauges allowing measurement between $10^{-3}$ and 1100 Pa (i.e., over the full range of the isotherms) with a resolution of 0.003% of full scale and a precision of 0.5% (in the range $10^{-3}$-1 Pa) and of 0.2% (in the range 1-1100 Pa) of the reading. A liquid $N_2$ bath temperature stabilization method was used to maintain a constant temperature (77.3 K) of the sample cell containing the adsorbent under study. The other part of the apparatus (including pressure gauges, inlet manifold, gas reservoir, etc, except the pumps) isolated from the cell by a valve, is stabilized at 30±0.5°C. Krypton is first introduced into this latter part and then into the adsorption cell by opening the valve. Upon adsorption, the cell pressure thus decreases and reaches a limiting value at equilibrium. Equilibration times ranged from 0.3 to 1 h per dose and were chosen experimentally. In several cases, we verified that there was no pressure value change 4 hours after reaching a limiting value. Since the pressure gauges are kept at a higher temperature than the sample vessel, the vapour pressure measurement can be influenced by thermal transpiration, especially in the low pressure region. So, the raw isotherms were corrected for the thermal transpiration effect (Takaishi et al. 1963).

Our experimental setup was optimized so that small mass samples - but with large specific surface areas - could indeed be examined. The sample mass was typically 10 mg. This value allowed precise study of the very first stages of Kr adsorption using the very sensitive capacitance manometers mentioned above. So, this value of sample mass allows introducing a correspondingly small quantity of gas and recording with maximal accuracy the pressure change during adsorption.

The sample was initially outgassed outside the adsorption chamber for 7 days at 100°C to a pressure lower than $10^{-4}$ Pa. After installation in the adsorption set-up, it was additionally outgassed for 1 day at 100°C to a pressure lower than $5.10^{-5}$ Pa. Further isotherm measurements always started with outgassing of the sample at room temperature (20°C) for 16 hours to a pressure of $< 5.10^{-5}$ Pa. Krypton ($\geq$99.998%, Fluka Analytical) was purified by pumping on the condensed phase inside the apparatus. The experiments were carried out with use of two adsorptive dosing (AD) protocols. These are designated IAD if the dose is increased from one injection to the next or CAD if the dose is constant.

It should be added that the void volume of the cell was determined without the sample in order to avoid any artifact of "helium entrapment" that could give a S-shaped form to the adsorption isotherms in the ultralow-pressure range, especially in the case of an insufficient equilibrium time. The volume of the cell was 32 cm$^3$ whereas that of the sample was 0.13 cm$^3$ so the sample volume is only 0.4% of the total cell volume.

Our experimental setup is manual not automatic. So, for an experiment that required several days, to continue to draw the isotherm on day n+1, after the outgassing indicated above the dosing for the first point of this day was that of the total quantity of Kr introduced into the cell during day n.

**2.3 Electron microscopy**

Samples were dispersed in absolute ethanol then deposited on a holey carbon film supported by a 300 mesh copper grid. Some of the high-resolution transmission electron microscopy (HRTEM) observations were done using a Philips CM200 operating at 200 kV. The others, HRTEM and scanning transmission electron microscopy (STEM) images, were done with a JEOL ARM 200F equipped with a cold field-emission gun and a probe Cs-corrector (correction of the spherical aberration). The microscope was operated at 80 kV to avoid damaging the SWCNTs. High-angle annular dark-field (HAADF) and bright-field (BF) STEM images were simultaneously obtained for collection semi angles of 45-180 mrad and 0-3.8 mrad, respectively, and a pixel time of 40 μs (1024x1024 pixels). For STEM imaging, the operating conditions allowed attaining a resolution close to 0.1 nm for both detection modes (measured on dispersed gold nanoparticles). The chirality of isolated SWCNTs was deduced from the fast Fourier transform (FFT) of the STEM BF images.

**2.4 Diffuse reflectance IR Fourier-transform (DRIFT) spectroscopy**

Infrared spectra of the samples were obtained using a Nicolet 8700 spectrometer equipped with a mercury cadmium telluride (MCT) detector and a KBr beam splitter. Infrared Powder Diffuse Reflectance data were collected with a Harrick Praying Mantis$^{TM}$ attachment and a high temperature reaction chamber. Spectroscopic grade KBr was used as background. The diffuse reflectance $R_s$ of the sample and $R_r$ of potassium bromide, used as a non-absorbing reference powder, were measured under the same conditions. The sample was prepared by mixing the SWCNTs with KBr (SWCNT fraction of 0.001), without compaction. The SWCNT reflectance is defined as $R=R_s/R_r$. The spectra are shown on a pseudo-absorbance (-logR) scale.

**3. Results and discussion**

**3.1 Structural characterization of SWCNT bundles**

Detailed TEM/STEM examinations of the sample show that it is quite free of any catalyst residue. There are a few carbon impurities. The SWCNT diameter range is somewhat larger than was claimed by NanoIntegris, 0.7-1.6 nm. Fig. 1 illustrates typical micrographs of the sample from low to high magnification.

Fig. 1 (a) presents an overall view. The SWCNTs combine into long bundles between which there are a few scattered metallic particles (arrow) as well as some hollow carbon shells of 4-5 nm in diameter (circle) (Fig. 1 (b)). It was estimated that on average there are 25-30 tubes per bundle. There is also a significant number of isolated nanotubes and very small bundles of 2 or 3 tubes (Fig. 1 (b-e)). The CNTs are quite straight over lengths of many tens of nm and appear to form loosely bound bundles. The box on Fig. 1 (b) illustrates three tubes all with closed ends, as was the case for most of the tubes observed in this study. Detailed STEM examination established a second type of carbon impurities on the surface of the tubes: small hollow spherical or ellipsoid structures of ~0.7-1.2 nm in size (Fig. 1 (d) and (e)). These hollow carbon "particles" are composed of one layer of carbon and might be fullerenes. The analysis of HRSTEM images shows that the distance between adjacent tubes can be variable. As seen on the BF-STEM image (Fig. 1 (d)), the interwall distance between the two tubes of 0.75 and 1.03 nm in diameter is between 0.34 nm and 0.70 nm. This could be explained by the presence of defects in the SWCNT structure along one (or both) of the tubes. It can also be seen that some hollow carbon particles lie between two tubes, locally increasing the distance between them. Whatever the reason underlying such a variable inter-tube distance, this observation is of particular interest because the regions of high dilation might play a role as additional entry ports to the interstitial channels of a bundle.

Fig. 1 (f) is an image of one part of a bundle which is perpendicular to the grid surface. The bundle is constituted of very different carbon nanotubes: the ratio of largest to smallest nanotube diameters is about two. Of particular interest here are the morphing and size of the interstitial channels separating the SWCNTs. Triangular-shaped channels defined by three "closely-packed" tubes are observed but there are other bigger channels (A to D) defined by four tubes (Fig. 1 (f)). Furthermore, the big tubes in the bundles are slightly deformed as noted for tube 2 on Fig. 1 (f). If other tubes are deformed, the flattering value is within the limits of resolution and it could not be estimated.

The microstructural observations allow concluding that the SWCNTs combine into small bundles. The SWCNT diameters are heterogeneous with a consequential range of interstitial channel shapes and sizes, some of which are large. The distance between adjacent tubes is highly variable. The majority of SWCNTs are closed. The average diameter of our nanotubes of ~1.2 nm allows for two-linear adsorption lines of krypton on the internal surface if the nanotubes are opened. Indeed, for the nanotubes of 1.2 nm means that there is an empty space in the tube center ~0.86 nm wide. The van der Waals diameter of Kr is 0.404 nm, which would allow the formation of a one- or two- linear structures as well as the layering transition. The structures formed on 1D or quasi-1D sites could have linear or zigzag arrangements (Jakubek et Simard 2004; Jakubek et Simard 2005; Arab et al. 2007; Hodak et al. 2003).

The SWCNT bundles examined in this study consisted of very different CNTs, containing bigger channels like channels A-D on Fig. 1 (f). Based on the HRTEM images the size is ~0.8x0.44 nm for channel A, ~1.0x 0.5 nm for channel B, ~0.6x0.9 nm for channel C and ~0.9x0.8 nm for channel D. So here, the formation of two-linear structures of adsorbed Kr is possible. However, the bundles structure are not fixed along their length: the bundles curve, the tubes can comprise defects in their structure or they can be slightly deformed. Furthermore, impurities like hollow carbon particles (Fig. 1 (d)) or metallic catalyst particles can be found between some tubes in the bundles. These can lead to a variable inter-tube distance and consequently to a various sizes and geometries of IC-sites. These IC-sites can thus be presented as <u>subnanoscale pore channels with alternating enlargements (voids) and constrictions (necks) along their axes</u>. The literature provides some interesting works on molecular simulation of adsorption in the interconnected mesopores (>2 nm) of a variety of shapes, sizes and surface characteristics (Maddox et al 1995; Horikawa T. et al. 2011; Sadeghi et al. 2012; Cimino et al. 2013; Puibasset 2014; Nguyen 2013). During gas adsorption the pores can exist in two states: first, unfilled, when the pore walls are covered by an adsorbed film but the pore center is occupied by vapor-like adsorbate, and secondly, filled, when the whole pore volume is occupied by a liquid-like, condensed adsorbate. The transition between these states is associated with the capillary condensation and capillary evaporation (desorption) transitions (Cimino et al. 2013; Horikawa T. 2011). When the pore is not blocked and the fluid has an interface with the vapor phase, evaporation occurs at an equilibrium relative pressure which is determined by the pore size. If the fluid is blocked by narrower pores, it cannot evaporate and hence becomes metastable. The formation inside the pores of bridge-like or large meniscus bridging structures (Puibasset 2014; Horikawa T. et al. 2011) characterize a metastable adsorption. The number of the metastable phases inside pore depends on the number of pore heterogeneities. The pore of ~1 nm and smaller (example: buckytubes) is too small to exhibit phase transitions such as capillary condensation (Maddox et al. 1995). However, one- or two-linear arrangements can have a low-density (vapor-like adsorbate) or a high-density (liquid-like) state of the structure. Accordingly, a layering transition from a low- to a higher-density state of the structure of adsorbed gas can occur. The liquid-like high-density structures can block the pore. Moreover, a very small subnanometer diameter pores with alternating enlargements and constrictions can be easily blocked by metastable "neck" structures.

## 3.2 Volumetric measurements

Two data sets with the two above-mentioned protocols were acquired for the 77 K isotherms. In the first measurement using the IAD protocol, a full isotherm to 167 Pa was drawn (Fig. 2(a)). Up to 0.05 Pa the Kr was added in doses from 0.06 to 0.157 mmol/g (mmol of Kr at standard temperature and pressure per g of sample); from 0.05 Pa to 3 Pa the doses were from 0.094 to 0.188 mmol/g; from 3 Pa to 20 Pa doses from 0.226 to 0.377 mmol/g, and from 20 Pa up to 167 Pa - 0.377 to 2.26 mmol/g. In a second series of measurements we studied more precisely the first part of the isotherm, up to 0.05 Pa (Fig. 3). Here, the CAD protocol was used. The second data set was realized with 11 different adsorptive doses, from 0.083 mmol/g to 1.200 mmol/g. As one can see from Fig. 3, the curves in the low-pressure region (LPR) obtained with the different AD steps, are not superposed. We will refer to these hereafter as the "branches" of the isotherm.

Before looking at the details, we first note that the isotherm obtained over the full range of pressure comprises two inclined steps. The second step extends from near 0.3 Pa to about 3 Pa (Fig. 2). Its position is independent of the protocol used and of the chosen AD value. This second step has the generally-observed structure for a rare gas adsorption isotherm on SWCNTs (Babaa et al. 2003; Arab et al. 2006; Muris et al. 2000). This is not at all true for the first step located in the LPR between about 0.0007 Pa and 0.05 Pa (Fig. 2 and 3) for which there are some intriguing features: the step behavior depends on both the chosen protocol and on the AD value. A pressure drop with increasing AD is noted using the IAD protocol (Fig. 2 (b)) and there are two small vertical substeps. In the measurements using the CAD protocol, the exact positions of the branches and small substeps depend on the value of AD (Fig. 3).

The measurements obtained with the IAD protocol in the LPR, are represented on Fig. 2 (b). There are two segments of the LPR isotherm each the result of a one-day experiment. Upon increasing the adsorptive dosing from 0.06 to 0.157 mmol/g the equilibrium pressure decreases with increasing AD and a small substep at 680 µPa is observed. The second segment was carried out on the day following the first experiment. The dosing for the first point was 0.938 mmol/g, i.e., the total quantity of Kr introduced in the cell during the first segment measurement. But as one can see the first point (FP) of the second measurement and the last point (LP) of the first one do not superpose. The equilibrium pressure of FP is lower than that of the LP for the same quantity of adsorbed Kr. The further points of the second segment were obtained with adsorptive dosing increased step-by-step from 0.087 to 0.147 mmol/g. There are no further drops in the equilibrium pressure as were observed in the first segment. There is a small substep at 2050 µPa.

We suppose that the observed adsorption dosing dependent structure of the low-pressure region of the isotherm is linked to adsorption mainly in the IC-sites since the majority of the SWCNTs are closed. So, even if there exists a small fraction of opened SWCNTs, they cannot define the character of the isotherm. Near the G-sites there can also exist one- or two-dimensional phases of adsorbed gas: the work of Migone's group (Talapatra et al. 2001; Talapatra et al. 2002) has shown for Ne, $CH_4$, Ar and Xe, that after the grooves are filled, some "stripe" phases may appear before the formation of the second layer on the external surface of bundles. But this manifests itself by the presence of substeps on the second step of the adsorption isotherm.

Let us look more closely at the obtained results. As one can see from Fig. 2 (b), as AD increases from 0.006 to 0.045 mmol/g the equilibrium pressure decreases from 4.12 mPa to 0.68 mPa. This could be explained by the formation of metastable adsorbed phases inside the IC-sites or by intrapore blocking. Fluid-like structures adsorbed in microporous (pore size below 2 nm) materials, due to a strong fluid-wall forces in the very narrow pores have properties that are very different from those of the usual fluid (Maddox 1995). While the bulk melting point of Kr is 115.78 K under ambient pressure, the melting point of Kr adsorbed in the subnanometer sized pores of SWCNTs bundles will be higher. Consequently, the intra-pore diffusion will be highly limited. The formation of specific pore neck structures can block the diffusion inside the pores. Moreover, the neck constrictions in the narrower pores separate the Kr which has diffused more deeply into the pores from the vapor phase, i.e. it leads to metastable adsorption. A subsequent higher pressure AD can remove the diffusion restriction and thus change the adsorption amount and the apparent equilibrium pressure.

Indeed, the first portion of Kr is small, 0.006 mmol/g. Part is adsorbed on the most attractive sites leading to a certain value of equilibrium pressure (P) in the cell. The next introduced portion of gas is a little greater (0.008 mmol/g) with the pressure ($P_{AD}$) of 305 times that of the equilibrium pressure in the cell after the first AD. This sudden pressure increase plays the role of a "piston": the newly arrived Kr adsorbs again on the most stable sites and "pushes" more deeply the already present Kr atoms. If the deeper penetration involves going beyond some constrictions, then we can imagine that some of the adsorbed Kr blocks the pore by formation of liquid-like structure in the pore neck. As a result, the apparent equilibrium pressure in the cell after this second AD is even lower than the previous. The pressure of the third portion of AD (0.017 mmol/g) is 612 times higher than the equilibrium pressure in the cell of the previous step. The Kr undergoes further "push". This process continues up to the point at which the pressure of the new AD is insufficient to further push the confined Kr structure. At this point, the sites are truly filled and no more Kr can be adsorbed here. On the adsorption curve this manifests itself by an equilibrium pressure increase. As the equilibrium pressure in the cell rises, the influence of the pressure of any newly introduced AD of Kr decreases since the smaller difference between P and $P_{AD}$ diminishes so the "piston effect" is less efficient. Another reason is in increased population of sites. This is probably why we can observe the inversion of the isotherm points only in a very low-pressure region when the amount of introduced and adsorbed Kr is small.

The measurements in the LPR obtained using the CAD protocol are presented on Fig. 3. It can be seen that the isotherm here has a multi-branch structure. Each branch is the result of a one-day experiment. The equilibrium pressures depend on the value of the adsorption dose. The points obtained for the four smallest AD values used are shown on Fig. 3 (a). On that obtained with the smallest AD (0.083 mmol/g), four substeps 1-4 are noted, at 680, 2050, 3080 and 4820 µPa. When the AD increased to 0.155 mmol/g, there are only two substeps at 3080 and 4820 µPa and their height is greater than in the case of the first branch. The third branch (0.172 mmol/g) is shifted to the right of the two preceding ones, i.e., to a higher equilibrium pressure. It has a substep at 4820 µPa. Increasing AD to 0.192 mmol/g leads to an even greater shift to the right and two new substeps (5 and 6) are observed, at 9740 and 15810 µPa. Further increasing AD up to 0.209 µPa leads to an unexpected downshift on the equilibrium pressure as shown in Fig. 3 (b). However, with respect to the substep obtained at a smaller value of AD, the height of the substep at 680 µPa is greater. Further increasing the AD from 0.209 to 0.560 mmol/g again progressively shifts the equilibrium pressures to the right (Fig. 3 (c)). Increasing AD to 1.2 mmol/g results again in a downshift (Fig. 3 (c)). We conclude that increasing AD influences the isotherm in a somewhat cyclic manner. Indeed, increasing AD first results in a corresponding progressive shift of adsorption branches to the right. Then at some higher AD, the new equilibrium pressure shifts to a lower value than the previous and a new branch is created to the left of the previous one. Examination of the LPR of the isotherm with an AD value above 1.2 mmol/g is impossible because any introduction of a greater quantity of Kr leads to saturation of all accessible 1D-sites. As a result, subsequent points of the isotherm correspond to the end of the LPR or the beginning of the second part of the isotherm.

The filling of pore can be accompanied by the formation of the metastable phases like it was described elsewhere (Puibasset 2014). While the first step on the isotherm represents filling of 1D sites, the occurrence of the substeps at different equilibrium pressures is attributed to various phases with different arrangements of the atoms. The position of the substeps is defined by the binding energy of the sites where this arrangement is realized. Let us look at Fig. 3 (b) on which there are five branches obtained with five different values of AD. The horizontal dashed line is arbitrarily drawn through the value of adsorbed Kr ~0.6 mmol/g. The line crosses the branches obtained with AD values of 0.209, 0.083, 0.155, 0.172 and 0.192 mmol/g at the points situated in the first, third, third, fourth and fifth substeps, respectively. These different points represent adsorption of the same quantity of Kr, but for each of these points the equilibrium pressure in the cell is different and depends on the AD value used. This is plausible if different confined phases of Kr with different densities or arrangements are formed. Wherever the way of adsorption (branch), all branches achieve the same amount of adsorbed Kr at the high pressure limit of the first step that corresponds to the filled sites. In other words, the adsorption inside the pores can pass through the formation of different intermediate metastable phases in the pores with different diameter. But finally, all pores are filled by adsorbate.

In the CAD protocol measurements, for a given value of AD there is no drop in equilibrium pressure as was the case for the IAD protocol. However, the smallest portion of Kr used in the CAD is nearly 14 times higher than the first portion of Kr in the IAD measurements. This means that the switch-back type behavior of the isotherm stemmed from metastable adsorption; it can be observed only through use of very small AD's and only at the very beginning of the LPR of the isotherm when the pore channels are almost empty or only slightly filled. Indeed, a similar pressure drop with increasing AD and the formation of a multi-branch structure in the LPR of the isotherm has been described (Jakubek and Simard 2004; Jakubek and Simard 2005) for Ar and Kr adsorption on opened SWCNTs. For the case of Ar, these authors used adsorbate dosing increased step-by-step (Jakubek and Simard 2004). For Kr adsorption the constant adsorption dosing protocol was used

(Jakubek and Simard 2005) with a very small AD (~0.07 cm$^3$ of Kr/g of adsorbent at standard temperature and pressure). Probably, when inside the pore channels there was a few adsorbate, the pores can be blocked by formation of liquid-like structure in the pore neck. Moreover, the adsorbate atoms are fractioned inside the pores forming the a liquid-like meniscus located mainly in the strongly attractive domains of the pore. In the IAD protocol when the pores contain a significant quantity of adsorbate atoms or in the CAD protocol with a large enough AD value, Kr atoms are distributed widely both in the pore necks and in the pore enlargements. The pores are not blocked. A certain partitioning of confined Kr inside the pores with formation of separated metastable phases is possible here, but due to the large population of sites it will not lead to pressure drop with increasing AD.

It is remarkable that the value of equilibrium pressure of the first two substeps is the same for the two used protocols. In addition, in the CAD protocol measurements the positions of certain substeps on different branches are the same but not their heights (e.g., substep 1 of the branches obtained with AD values of 0.083 and 0.209 mmol/g (Fig. 3 (a-b)). The position of a substep on the adsorption isotherm is defined by its binding energy (Arab et al. 2007). The substep height is linked to the population of the sites. The results thus suggest a certain selectivity or a preferential metastable phase formation according to the AD value. There appear to be a certain number of accessible adsorptive sites in which one given adsorbate atomic arrangement can be made. The site filling and the adsorbate structure/density depend on the value of the AD as well as on the AD protocol.

So, we hypothesize that the multi-branch structure and branch switching that we have observed may be explained by the existence of some metastable phases of confined Kr and that a transition from a low-density to a high-density state occurs in the IC-sites.

To examine the possibility of modifying the structural characteristics of CNTs during volumetric measurements or of opening the SWCNTs, we verified the reproducibility of our measurements. Points from two different measurements with the same adsorptive dosing (0.39 mmol/g) are presented on Fig. 4. The second of these two measurements was carried out one month after the first and between the two, several other adsorption experiments on the same sample were done with different AD values. As can be seen, the results are reproducible. We conclude that even if there exist certain environmental or experimental influences not fully taken into account during our experiments they do not perceptibly affect the characteristics of the isotherm. Moreover, we have carried out the adsorption isotherms for three samplings of the HiPco SuperPureTubes™. The isotherms of all three samplings are absolutely identical which confirms good sample homogeneity and reproducibility of the observed phenomenon. In addition, detailed electron microscopy examinations of the samples after the volumetric measurements did not establish any changing of the structural characteristics.

To verify that the observed adsorption behavior in the LPR is linked to the formation of pore blocking metastable phases <u>inside the IC-sites</u>, we have studied the adsorption of the same HiPco SWCNTs after an $H_2SO_4$ treatment. Indeed, Kuznetsova et al. (2000) showed that upon such an acid treatment, the grafted functional groups such as quinone or carboxylic acid block the entry ports of the nanotubes. The very soft conditions of our acid treatment were chosen to avoid the possible cutting or significant opening of SWCNTs, i.e. to preserve the bundle morphology and to graft the functional groups at the sites of the entry ports of IC- and T-pores. It is known (Feng et al. 2004) that the most reactive sites on the SWCNT bundles are the ends, the bundles curves and the connecting regions between different bundles, i.e. all possible IC entry ports. So, the soft oxidation can allow grafting here of the oxygen-containing functional groups that will block the adsorption inside the pore sites of SWCNT bundles.

A typical DRIFT spectrum of $H_2SO_4$-treated SWCNTs, is shown on Fig. 5. There are two bands characteristic of carboxylic acid (at 1730 $cm^{-1}$) and quinone (at 1660 $cm^{-1}$) groups. A very broad band with a maximum near 3250 $cm^{-1}$ can be identified with -OH from the carboxylic acid or the phenol groups attached to SWCNTs. There is also a triplet at 2900 $cm^{-1}$ (2960, 2925 and 2850 $cm^{-1}$) that is identified with $C-H_n$ functional groups (Kim et al. 2005). A very small broad band of hydroxyl groups as well as a low intensity triplet at 2900 $cm^{-1}$ are observed on the curve of pure SWCNTs. These functions probably come from the purification procedure that often involves acid treatments to remove metal catalyst impurities from HiPco samples.

During the soft acid treatment used the electrophilic reagent first attacks some C=C bonds producing hydroxyl groups which in turn can be converted into quinone groups, and finally to carboxylic acid groups (Zhang et al. 2003). The quinone groups may be considered as intermediates in the oxidation process (Zhang et al. 2003). The ratio between the peaks at 1730 $cm^{-1}$ and 1660 $cm^{-1}$ corresponds to the ratio between carboxylic and quinone groups. As the oxidation proceeds, this ratio increases. For our oxidized sample the ratio between carboxylic acid and quinone groups is approximately the same. TEM examination of the ox-SWCNT shows that indeed there is no change in morphology of the sample, i.e., the SWCNTs have not been cut nor is there any significant SWCNT opening. So, the TEM results corroborate the conclusion of the DRIFT analysis: even if some carboxylic acid and quinone groups were grafted on the SWCNT surface, the number is insufficient to change the bundle morphology.

Let us now examine the Kr adsorption isotherm of ox-SWCNTs (Fig. 6): it is significantly different from that of the pure-SWCNTs as shown above. There are no branches or equilibrium pressure drops with either the CAD or the IAD protocols: in fact, the shape of the adsorption curve is independent of the protocol used and of the chosen AD value. The curve starts directly by the first step which extends from about 0.005 to about 0.02 Pa, i.e. it is much narrower than in the case of untreated SWCNTs. The position of the second step coincides with that of pure-SWCNT sample.

As discussed above, the oxidation was chemically "soft" with no tube cutting, creation of few defects and no change in bundle morphology. We suggest that the functional groups are located mainly on the SWCNT ends thereby blocking the entry ports of the ICs similar to the blocking of the tube inner channels as proposed by Kuznetsova et al. (2000). Such IC-blocking upon oxidation on bundles which are otherwise very similar to our raw bundles, thus strongly supports the our interpretation of the role of the open channels during adsorption.

More extensive experimental investigations in a broad range of temperatures and with other gases such as Ne with a van der Waals diameter of 0.308 nm or Xe with a diameter of 0.432 nm as well as advanced theoretical simulations are necessary to fully understand adsorption in the IC-sites and the formation of various blocking metastable phases with different arrangements of Kr. Additional experiments will be carried out.

## 4. Summary and conclusions

In this work we have carefully studied krypton adsorption at 77K on a sample of HiPco SuperPureTubes™. Most of the SWCNTs in the bundles are closed, on average there are 25-30 per bundle and they are heterogeneous in diameter with a consequential range of IC sizes, some of which are large. Moreover, the inter-tube distances are variable, also affecting the size and the geometry of the IC-sites. Such sites in SWCNT bundles can be presented as subnanoscale pore channels with alternating enlargements (voids) and constrictions (necks) along the tube axes.

Two different volumetric adsorption protocols were used, one with an increasing Kr dose per injection, one with a constant dose. We have established adsorption dosing dependent structures of the low-pressure region of the isotherm. To eliminate any possibility of artifacts or that the volumetric protocol itself might modify the CNT structure, studies on different samplings and reproducibility studies using identical protocols confirmed that the results are indeed reproducible. Detailed electron microscopy examination of the samples after the volumetric measurements strengthened the argument that the volumetric experiments did not change the structural characteristics of the samples.

The new adsorption features include the existence of AD-dependent behavior comprising branches, substeps and branch switching. An equilibrium pressure drop with increasing AD is noted using the

IAD protocol which we explain by the formation of metastable adsorbed phases inside the IC-sites or by intrapore blocking.

In the CAD protocol measurements six and in the IAD protocol measurements two well-pronounced substeps corresponding to the formation of various phases of confined Kr were established. The height of the same substep obtained in different measurements depends on the AD value which can strongly influence the population of the site. In the measurements using the CAD protocol, the exact position of the first step is AD-dependent. The increasing AD influences the isotherm in a cyclic manner. Whatever the scenario of the pore filling, all branches achieve the same value of adsorbed Kr at the high pressure limit of the first step of the isotherm that corresponds to the filled pore sites.

Finally, Infrared studies on an acid-treated sample show the existence of carboxylic acid and quinone groups on the nanotube surface. These groups block the entry ports for adsorption on the ICs surface. These lead to a Kr adsorption isotherm the shape of which is independent of the protocol used and of the chosen AD value and thus lend support to the important role of IC sites in the raw samples.

Further experiments are in progress to shed more light on the complex interactions between adsorbate and substrate in these heterogeneous systems.

## ACKNOWLEDGMENTS

Acknowledgment is made to the French Agence Nationale de la Recherche (ANR) (grant ANR-10-BLANC-0819-01-SPRINT) and the Région Lorraine (grant 30031172) for their support.

The authors thank Prof. C. Carteret, Dr. M. Dossot and Dr. S. Fontana for fruitful discussions and Dr. J.-F. Marêché for important technical assistance.

# - Figures -

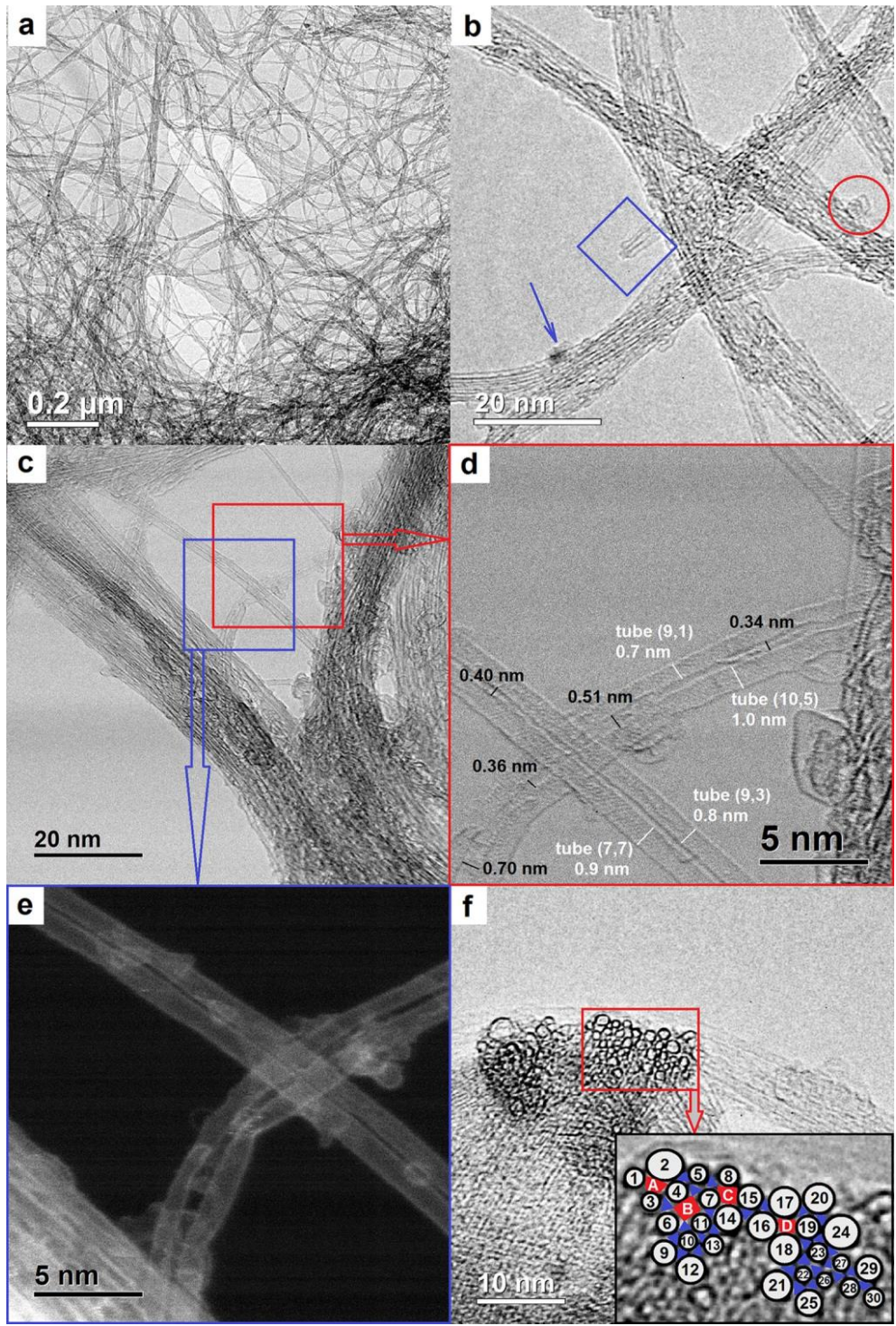

Fig. 1: TEM (a), HRTEM (b, f), BF-STEM (c, d) and HAADF (e) images of HiPco SuperPureTubes™ sample. The inset at the bottom of image (f) is a zoom of the presented bundle. The blue triangles indicate interstitial channels defined by three tubes; the letters A to D designate ICs defined by 4 tubes.

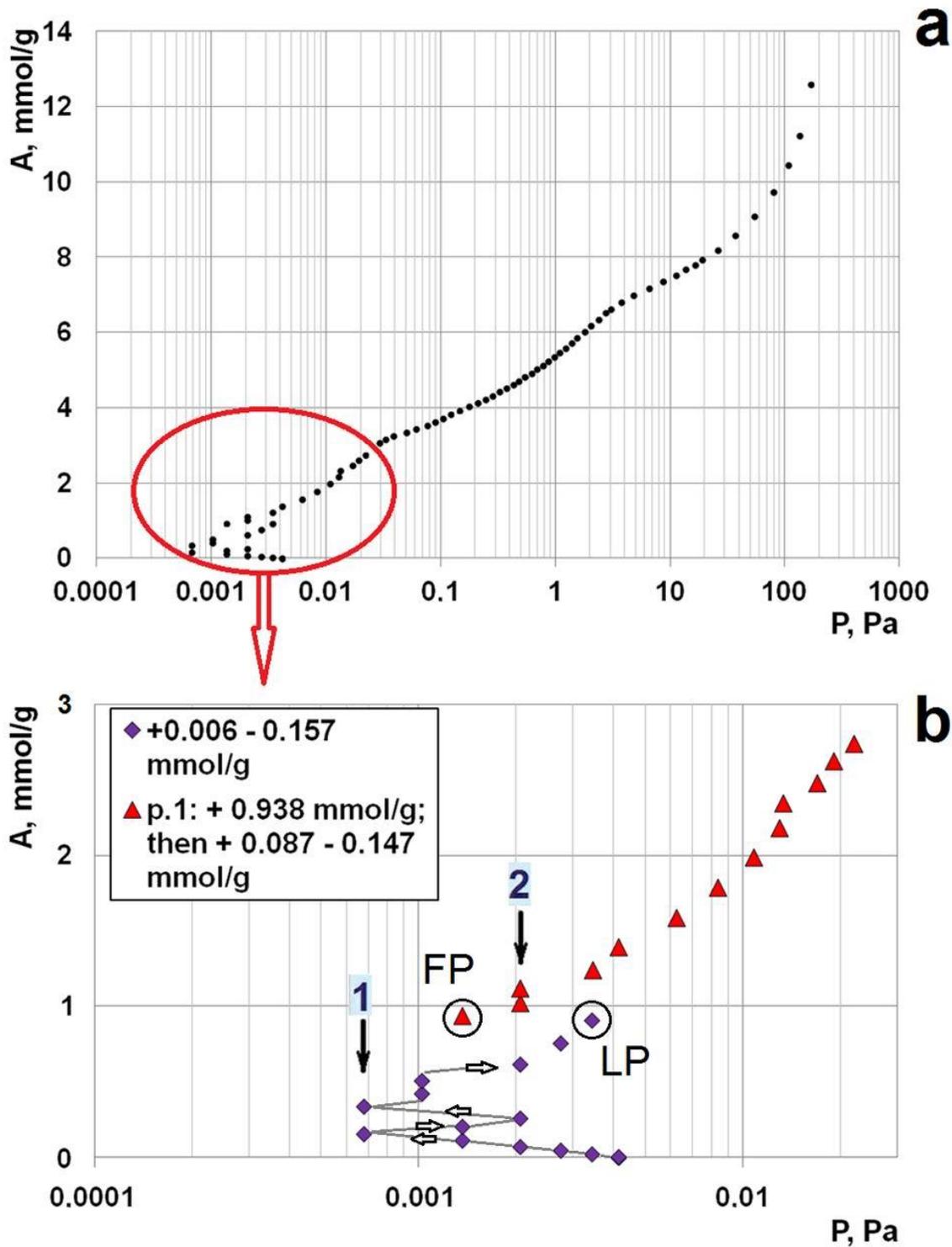

Fig. 2: (a) 77K isotherm of Kr adsorbed on HiPco SuperPureTubes™ sample. (b) Low-pressure region of the isotherm. The vertical arrows 1 and 2 designate the substeps at equilibrium pressures 680 and 2050 μPa. The first point (FP) of the second measurement and the last point (LP) of the first one are circled. From bottom to top the four horizontal arrows indicate the successive directions of equilibrium pressure evolution.

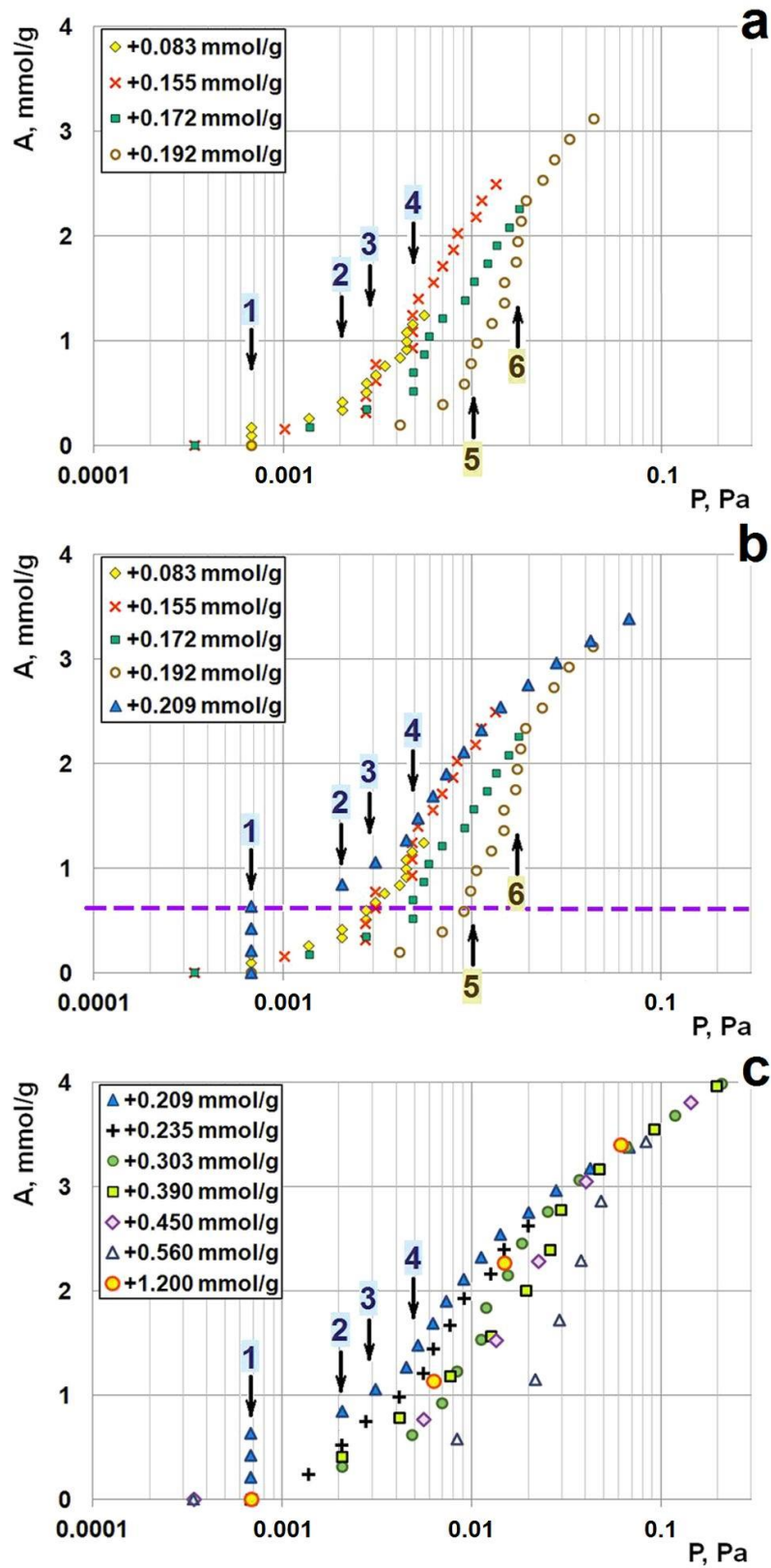

Fig. 3: (a-c) Low-pressure region of the 77K isotherm of Kr adsorbed on HiPco SuperPureTubes™ sample obtained with the 11 different constant adsorptive doses indicated. The arrows 1 to 6 designate the substeps at equilibrium pressures of 680, 2050, 3080, 4820, 9740 and 15810 µPa.

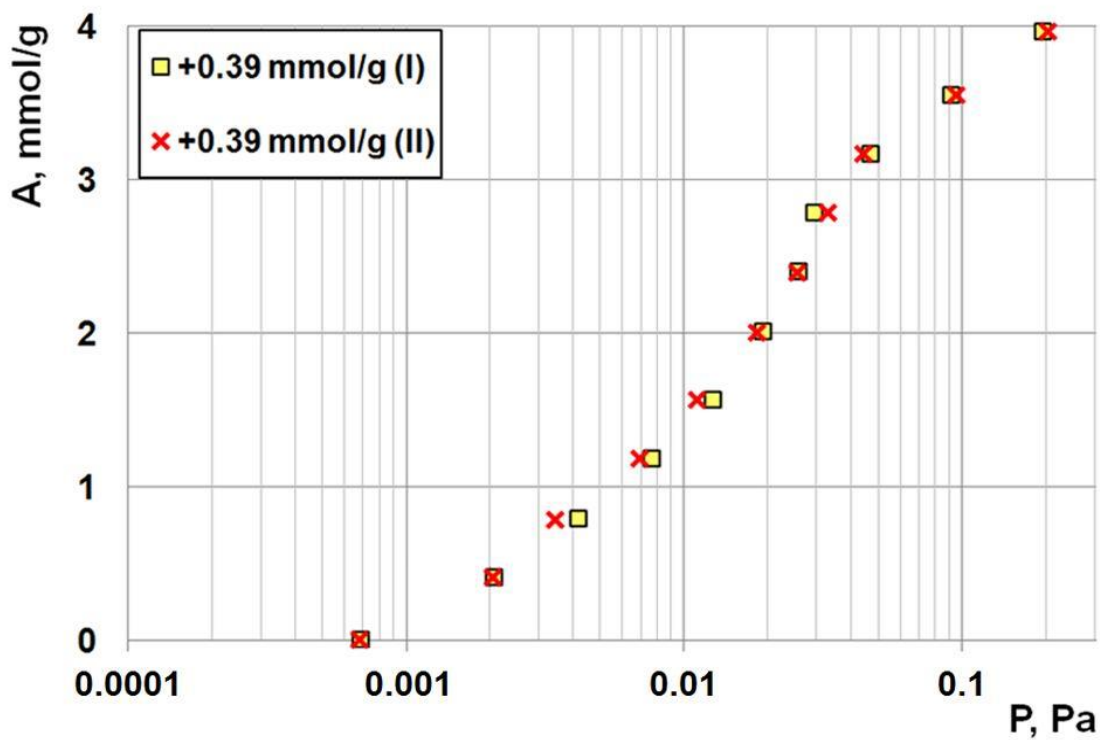

Fig. 4: Illustration of the good measurement reproducibility. Points from two different measurements with the same adsorptive dosing are indicated by squares and crosses.

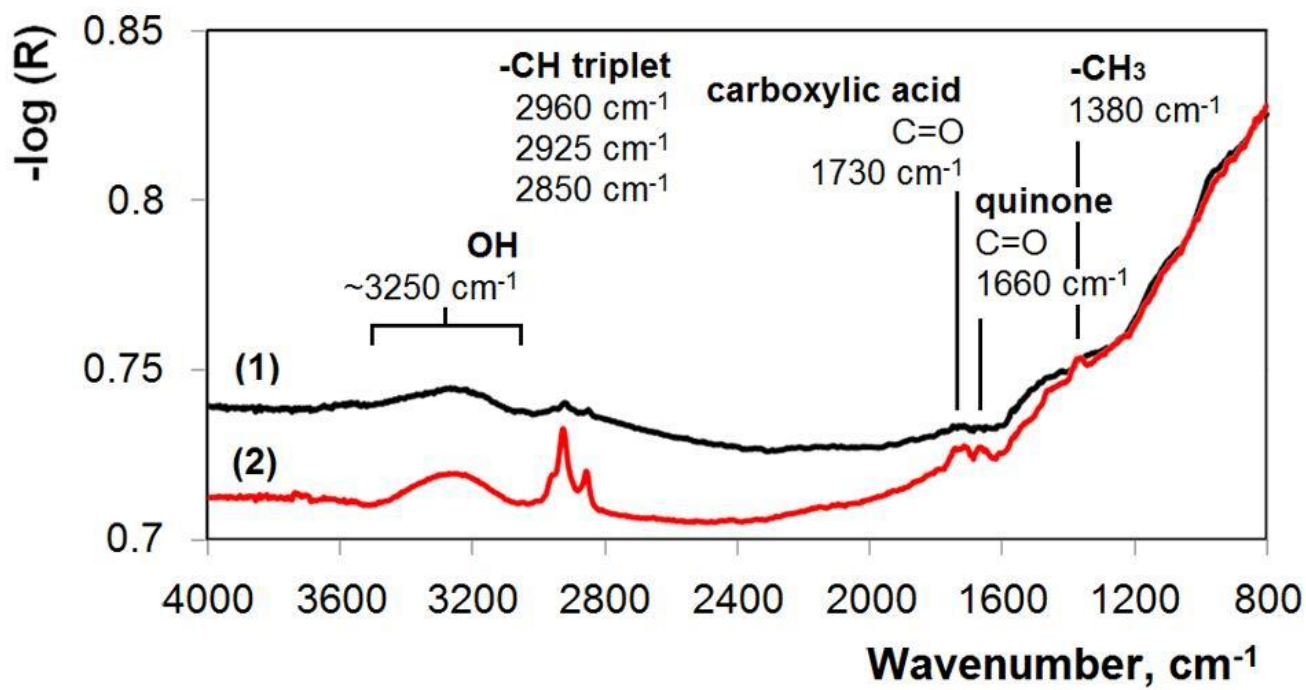

Fig. 5: DRIFT spectra of HiPco SuperPureTubes™ (1) and $H_2SO_4$-treated SWCNTs (2).

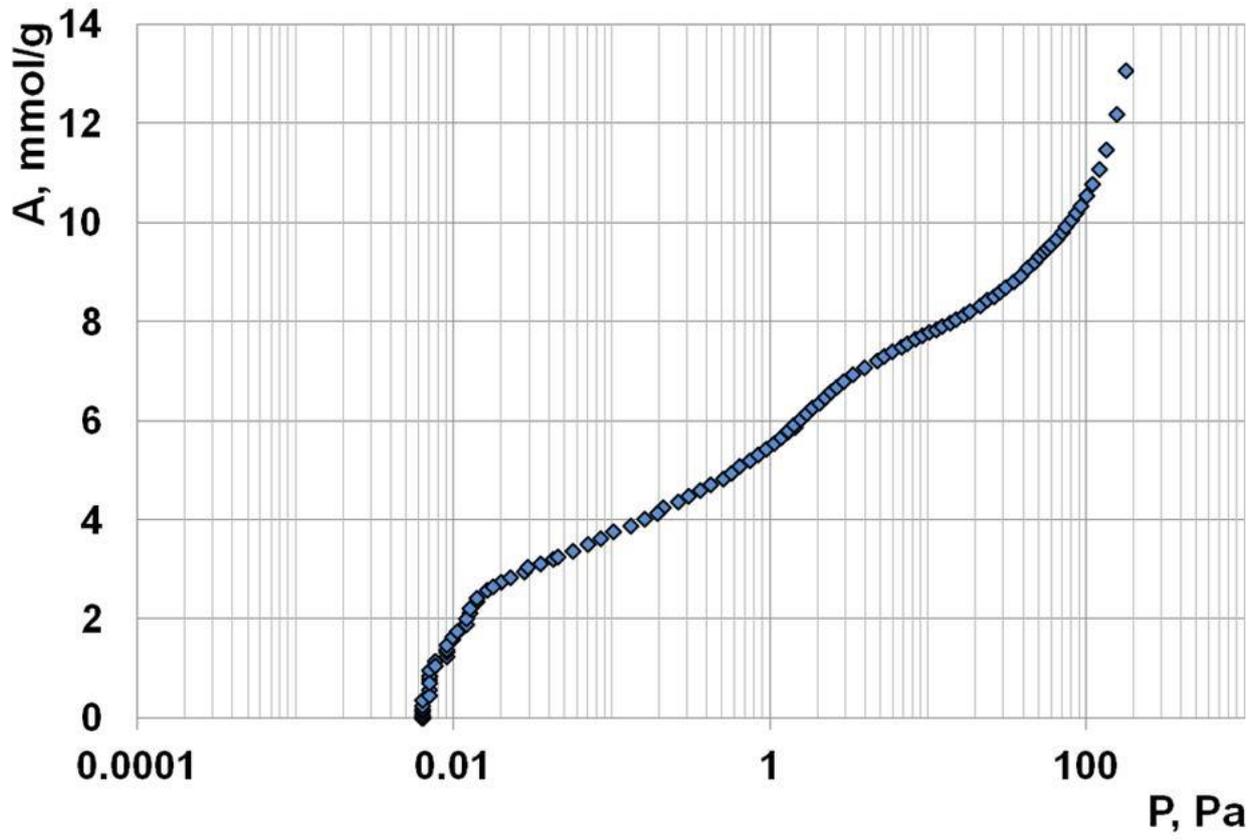

Fig. 6: 77K isotherm of Kr adsorbed on $H_2SO_4$ treated HiPco SuperPureTubes™ sample.